# SAMPLING ALMOST PERIODIC FUNCTIONS WITH RANDOM PROBES OF FINITE DENSITY

P.COLLET

Centre de Physique Théorique
Laboratoire CNRS UPR 14
Ecole Polytechnique
F-91128 Palaiseau Cedex (France)

**Abstract**: We consider the problem of reconstructing a function given its values on a set of points with finite density. We prove that with probability one, the values of an almost periodic function on a random array of points (with finite density) completely determine the function. We also give some properties of the associated Blaschke product.

# I. INTRODUCTION.

The study of dynamical systems in unbounded domains leads to many new and interesting questions. A typical (and important) example of such systems is the the complex Ginzburg-Landau equation

$$\partial_t A = (1 + i\alpha)\Delta A + A - (1 + i\beta)A|A|^2 \qquad (CGL)$$

where $\alpha$ and $\beta$ are two real parameters and $A(x,t)$ is a complex valued function of a spatial variable $x \in \mathbf{R}^n$ and time $t \in \mathbf{R}^+$. Although the results below will be of general nature and do not depend on a particular equation, it is interesting to have in mind a particular and important case which provides a natural setting for the problems. The Ginzburg-Landau equation occurs naturally in bifurcation of continuous spectra at non zero wave number (see [C.E], [V.H.] [K.S.M.]) and is often used as a paradigm for nonlinear dissipative evolutions in unbounded domains. This equation is well behaved in dimension 1 and 2 for all the (finite) values of the parameters, in particular bounded uniformly continuous initial conditions give rise to bounded solutions (even uniformly bounded and analytic in a uniform strip at large enough time, see [C1.], [C2.], [T.B.D.vH.T]). In [C.T.] we proved the existence of determining nodes and determining modes for CGL with finite spatial density. A natural question arising from these results is the determination of a solution $A$ given a discrete set of data. For example one could imagine experimentally a net of probes measuring at fixed positions the values of $A$ at a given time. One would then ask about the properties of $A$ that can be reconstructed with these data, in particular if one can reconstruct $A$ on all of $\mathbf{R}^n$ (we refer to [C.J.T.] for a related question in bounded domain). An important natural constraint on the set of probes is that it should have a finite density. This is of course physically reasonable, however this constraint destroys immediately any hope of identifying a general function (even very regular) due to the existence of the Blaschke product. If $(x_n)$ is a bi-infinite sequence of real numbers such that for a $\delta > 0$ the infinite product

$$\prod_{n \in \mathbf{Z}} \text{Sign}(x_n) \tanh(\pi x_n / 4\delta)$$

is convergent and non zero, then the infinite product

$$B(z) = \prod_{n \in \mathbf{Z}} \text{Sign}(-x_n) \tanh(\pi(z - x_n)/4\delta)$$

is convergent for any $z$ in the strip $|\Im(z)| \leq \delta$ and defines a non trivial analytic function in that strip with modulus bounded by one. We refer to [H.] or [G.] for the existence and properties of the Blaschke products. Of course this function is equal to zero at all the points $x_n$. Therefore, if $A$ is a function analytic and bounded in the strip $|\Im(z)| \leq \delta$, the function $A + B$ has the same property and is indistinguishable from $A$ if we only have observations at the points $x_n$.

One may however ask if some positive result can be obtained by restricting the set of functions to be observed. In particular the CGL equation has wave solutions, namely

$$A(x,t) = \sqrt{1 - q^2} e^{i(\omega t + qx)}$$

with $q \in [-1, 1]$ and $\omega = (\beta - \alpha)q^2 - \beta$. A natural question is now if we can find a sequence of probes $(x_n)$ with finite density such that any periodic function equal to zero at all these locations is guaranteed to be identically zero.

The easiest arrangement of probes is of course a periodic one, for example $x_n = n$. However for this arrangement the program of detecting even a periodic function fails immediately because if $f$ is a periodic function of period one with $f(0) = 0$ but otherwise non trivial, we have of course $f(x_n) = 0$ for all $n$, and the function $f$ escapes our observation.

This leads to the idea of using random locations for the probes (still maintaining a finite density). There are many point processes in $\mathbf{R}^n$ and we will consider below only particular ones. For simplicity of notations we will only consider the case $n = 1$, although theorem 1 below is valid in any dimension with obvious modifications and essentially the same proof.

We construct our first one dimensional point process $\mathcal{P}$ as follows. We first fix a positive length $\lambda$. We then consider a partition of the line by intervals of length $\lambda$. Then in each interval we choose a point at random with uniform distribution and we do this independently in each interval. This construction leads to a point process with average density $\lambda^{-1}$. We denote by $(x_n)$ ($n \in \mathbf{Z}$) the configurations of this point process. In other words, there is a bi-infinite sequence of i.i.d. random variables $(X_n)$ uniformly distributed on $[0, \lambda]$ such that

$$x_n = n\lambda + X_n \ .$$

**Theorem 1.** *Almost every configuration $(x_n)$ of the point process $\mathcal{P}$ has the property that if $f$ is any complex valued almost periodic function satisfying*

$$f(x_n) = 0 \quad \text{for all } n \in \mathbf{Z}$$

*then $f \equiv 0$.*

We recall that a function $f$ is almost periodic if it is uniformly continuous and if for any $\epsilon > 0$ there is a $\Lambda(\epsilon) > 0$ such that any interval $[a, a + \Lambda(\epsilon)]$ contains a number $\tau(a, \epsilon)$ such that

$$\sup_x |f(x + \tau(a, \epsilon)) - f(x)| \leq \epsilon \ .$$

We refer to [B.] or [F.] for the main properties of almost periodic functions.

Theorem 1 says that for almost every choice of the distribution of probes, almost periodic functions are determined by their values on the random array. This is a rather striking improvement over the case of regular arrays where as noticed above one can find periodic functions which are non trivial but equal to zero on the points of the array.

One may wonder how far one can go in the opposite direction, namely if the Blaschke product $B$ has some interesting properties. This is indeed the case.

**Theorem 2.** *Almost every configuration of $\mathcal{P}$ is such that the Blaschke product*

$$B(x) = \prod_{n \in \mathbf{Z}} \text{Sign}(-x_n) \tanh(x - x_n)$$

is non trivial, and belongs to the class $S'$ of Wiener.

Notice also that in the above theorem, $B(x)$ is analytic and bounded in a strip around the real axis.

We recall that $S$ is the set of bounded measurable functions $f$ such that the following limit exists for all $x \in \mathbf{R}$

$$C_{f,f}(x) = \lim_{L \to \infty} \frac{1}{2L} \int_{-L}^{L} \overline{f}(x+y) f(y) \, dy \ .$$

The set $S'$ is the subset of $S$ where the above limit depends continuously on $x$. This space of functions for which the spatial averaged correlations exist was introduced and studied by Wiener (see [W.] for more properties), and in particular these are functions which have a well defined power spectrum. It is also easy to verify using Birkhoff's ergodic theorem that if $\mu$ is a measure on the space of functions which is ergodic by space translation and supported by bounded functions, it is in fact supported by $S$. To be in $S'$ is therefore a natural requirement for a (regular) solution of a non linear evolution in unbounded domain. This assumption is in fact used implicitly very often in the Physical literature where averages are needed without an explicit knowledge of the averaging measure (see for example [R.B.S.] for recent numerical results and references).

In fact this argument extends to higher correlations. Namely, define $S^\infty$ to be the space of all bounded measurable functions $f$ such that all the limits

$$\lim_{L \to \infty} \frac{1}{2L} \int_{-L}^{L} dy \prod_{j=1}^{n} f^{\#_j}(x_j + y)$$

exist, and are continuous functions of $x_1, \cdots, x_n$, where $f^\# = f$ or $\overline{f}$. Then any translation invariant ergodic measure supported by uniformly continuous bounded functions is supported by $S^\infty$. It is not very difficult to prove that indeed $S^\infty$ is invariant by the CGL evolution equation (the subspace of almost periodic functions is also invariant).

The rest of the paper is organized as follows. Section II gives a proof of Theorem 1. Proving Theorem 1 for a given period or for a countable number of periods is easy. However the difficulty is that we have to deal with an uncountable number of periods. The proof requires a precise estimate of the probabilitry of some small sets (related to rational periods) which is obtained using ideas from the theory of large deviations. One then approximates irrational frequencies by rational ones in such a way that the error can be controlled using the previous estimates.

In section III we give a proof of Theorem 2 based on a law of large numbers for asymptotically weakly coupled random variables. Other versions of Theorem 1 using only fluctuations of regular arrays of points or Poisson point processes are briefly discussed in section IV.

## II. ALMOST PERIODIC SAMPLING.

In order to prove Theorem 1, note that by a simple rescaling it is enough to assume that $\lambda = 1$. We will need later on a large deviation estimate which is provided by the following Lemma.

**Lemma 3.** *For any $k \in \mathbf{N}$, any $m \in \mathbf{Z}$, any $p, q, r \in \mathbf{N}$ such that $0 \leq p < q$ (and $p > 0$ if $m = 0$), the set*

$$\mathcal{A}(k, m, q, r, p) = \left\{ \left| \frac{1}{(2r+1)q} \sum_{s=-r}^{r} \sum_{l=0}^{q-1} e^{2\pi i (m+p/q) x_{sq+l}} \right| \geq \frac{1}{k} \right\}$$

*satisfies*

$$\mathbb{P}(\mathcal{A}(k, m, q, r, p)) \leq 4 e^{-\frac{1}{512 k^2}(2r+1)q} \ .$$

**Proof.** Let $A(m, q, r, p)$ and $B(m, q, r, p)$ be the random variables defined by

$$A(m, q, r, p) = \sum_{s=-r}^{s=r} \sum_{l=0}^{l=q-1} \cos\left(2\pi(m + p/q) x_{sq+l}\right)$$

$$B(m, q, r, p) = \sum_{s=-r}^{s=r} \sum_{l=0}^{l=q-1} \sin\left(2\pi(m + p/q) x_{sq+l}\right) \ .$$

We are going to derive a large deviation estimate for these two random variables.

Let $\alpha = 1/64$, we have by the independence of the random variables $x_n$

$$\mathbb{E}\left(e^{\alpha k^{-1} A(m,q,r,p)}\right) = \prod_{s=-r}^{s=r} \prod_{l=0}^{l=q-1} \mathbb{E}\left(e^{\alpha k^{-1} \cos(2\pi(m+p/q) x_{sq+l})}\right) .$$

We now observe that for any real $z$ with $|z| < 1/2$ we have

$$|e^z - 1 - z| \leq 2|z|^2 \quad \text{and also} \quad |\log(1+z) - z| \leq 2|z|^2 \ .$$

If $S_{m,q,p,s,l}$ denotes the random variable

$$S_{m,q,p,s,l} = \cos\left(2\pi(m + p/q) x_{sq+l}\right)$$

we then have (since $|S_{m,q,p,s,l}| \leq 1$)

$$\left| \mathbb{E}\left(e^{\alpha k^{-1} S_{m,q,p,s,l}}\right) - 1 - \alpha k^{-1} \mathbb{E}(S_{m,q,p,s,l}) \right| \leq 2 \frac{\alpha^2}{k^2} ,$$

which implies since $k \geq 1$ and $4\alpha k^{-1} < 1$

$$\left| \log \mathbb{E}\left(e^{\alpha k^{-1} S_{m,q,p,s,l}}\right) - \alpha k^{-1} \mathbb{E}(S_{m,q,p,s,l}) \right| \leq 8 \frac{\alpha^2}{k^2} \ .$$

This implies immediately

$$\mathbb{E}\left(e^{\alpha k^{-1} S_{m,q,p,s,l}}\right) \leq e^{\alpha k^{-1} \mathbb{E}(S_{m,q,p,s,l}) + 8\frac{\alpha^2}{k^2}} .$$

We now observe that since

$$x_{sq+l} = sq + l + X_{sq+l}$$

and since the random variables $X_{sq+l}$ are independent, the number $\mathbb{E}(S_{m,q,p,s,l})$ is equal to

$$\mathbb{E}(\cos\left(2\pi(lp/q + (m + p/q)X)\right)) ,$$

where $X$ is uniformly distributed on $[0, \lambda]$. It now follows easily from this identity that if $p \neq 0$

$$\sum_{l=0}^{l=q-1} S_{m,q,p,s,l} = 0 .$$

If $p = 0$ and if $m \neq 0$, then by direct computation we have $\mathbb{E}(\cos(2\pi mX)) = 0$. This implies in all cases

$$\mathbb{E}\left(e^{\alpha k^{-1} A(m,q,r,p)}\right) \leq e^{8\alpha^2 k^{-2}(2r+1)q} .$$

Therefore

$$e^{\alpha k^{-2}(2r+1)q/4} \mathbb{P}\left\{A(m,q,r,p) > \frac{1}{4k}\right\} \leq \mathbb{E}\left(e^{\alpha k^{-1} A(m,q,r,p)}\right) \leq e^{8\alpha^2 k^{-2}(2r+1)q}$$

which implies

$$\mathbb{P}\left\{A(m,q,r,p) > \frac{1}{4k}\right\} \leq e^{-\frac{1}{512k^2}(2r+1)q} .$$

The same estimate holds for the random variables $-A$ and $\pm B$ and this proves the Lemma.

We now define a set of configurations of the random variables $(x_n)$ which is suitable for the sampling of periodic functions. For $k \in \mathbf{N}$ and $m \in \mathbf{Z}\setminus\{0\}$ let

$$\mathcal{B}(k, m) = \left\{ \begin{array}{c} \exists \xi \in \mathbf{N} \text{ such that } \forall r \in \mathbf{N} \text{ with } r > \xi, \ \forall q \in \mathbf{N} \text{ and } 0 \leq p < q \\ \left|\frac{1}{(2r+1)q} \sum_{s=-r,l=0}^{s=r,l=q-1} e^{2\pi i(m+p/q)x_{sq+l}}\right| \leq \frac{1}{k} \end{array} \right\}$$

Similarly, we define

$$\mathcal{B}(k, 0) = \left\{ \begin{array}{c} \exists \xi \in \mathbf{N} \text{ such that } \forall r \in \mathbf{N} \text{ with } r > \xi, \ \forall q \in \mathbf{N} \text{ and } 0 < p < q \\ \left|\frac{1}{(2r+1)q} \sum_{s=-r,l=0}^{s=r,l=q-1} e^{2\pi i(m+p/q)x_{sq+l}}\right| \leq \frac{1}{k} \end{array} \right\} .$$

And finally let
$$\mathcal{B} = \bigcap_{k \in \mathbf{N}\ m \in \mathbf{Z}} \mathcal{B}(k,m) \ .$$

**Corollary 4.** *The set $\mathcal{B}$ has probability 1.*

**Proof.** It is enough to prove that each set $\mathcal{B}(k,m)$ has probability 1, or equivalently that the complement has probability zero. On the other hand this complement satisfies
$$\mathcal{B}^c(k,m) \subset \bigcap_{w \in \mathbf{N}} \bigcup_{r \geq w} \bigcup_{q \in \mathbf{N}} \bigcup_{p < q} \mathcal{A}(k,m,q,r,p) \ ,$$
and the result follows at once from the estimate of Lemma 3, since (for $m \neq 0$)
$$\mathbb{P}\left( \bigcup_{r \geq w} \bigcup_{q \in \mathbf{N}} \bigcup_{p < q} \mathcal{A}(k,m,q,r,p) \right) \leq \sum_{r \geq w} \sum_{q \geq 1} \sum_{p=0}^{q-1} \mathbb{P}\left( \mathcal{A}(k,m,q,r,p) \right) \ ,$$
which tends to zero when $w$ tends to infinity, and similarly for $m = 0$.

**Proposition 5.** *Almost every configuration of $\mathcal{P}$ is such that for any real non zero number $\omega$, we have*
$$\lim_{L \to \infty} \frac{1}{(2L+1)} \sum_{-L \leq n \leq L} e^{2\pi i \omega x_n} = 0 \ .$$

*In particular, if $g$ is a trigonometric polynomial*
$$g(x) = \sum_{k=0}^{a} B_k e^{\omega_k x} \ ,$$
*with $B_k \in \mathbf{C}$, $\omega_k \in \mathbf{R}$, $\omega_0 = 0$, $\omega_k \neq 0$ for $k = 1, \cdots, a$, then*
$$\lim_{L \to \infty} \frac{1}{(2L+1)} \sum_{-L \leq l \leq L} g(x_l) = B_0$$

Note that this Proposition says in particular that the limit exists.

**Proof.** We first fix an integer $k > 0$. For a given non zero real $\omega$ we define $m$ by $m = [\omega]$ ($[\,\cdot\,]$ denotes the integer part).

If $\omega$ is irrational, let $(p_j, q_j) \in \mathbf{N}^{*2}$ ($p_j < q_j$) be the sequence of successive best rational approximations for $\omega - m$. We recall that the sequence $(q_j)$ is strictly increasing and for any $j \in \mathbf{N}$ we have (see [H.R.])
$$|\omega - m - p_j/q_j| \leq \frac{1}{q_j q_{j+1}} \ .$$

Since the sequence $q_j\sqrt{q_{j-1}}$ is increasing, for any $L \in \mathbf{N}$ large enough, we can find a unique integer $j > 1$ such that

$$q_j\sqrt{q_{j-1}} \leq L < q_{j+1}\sqrt{q_j} \,.$$

We then define an integer $r_j$ by

$$L = r_j q_j + t_j \quad \text{with} \quad 0 \leq t_j < q_j \,.$$

Note that from our choice of $j$ it follows that

$$\sqrt{q_{j-1}} - 1 \leq r_j \leq \frac{q_{j+1}}{\sqrt{q_j}} \,,$$

and by taking $L$ large enough, we can assume $q_{j-1} > 1$. We now observe that since $|L - r_j q_j| \leq q_j$, we have

$$\left| \frac{1}{(2L+1)} \sum_{-L \leq l \leq L} e^{2\pi i \omega x_l} - \frac{1}{(2r_j+1)q_j} \sum_{s=-r_j, l=0}^{s=r_j, l=q_j-1} e^{2\pi i \omega x_{sq_j+l}} \right| \leq \frac{4q_j}{2L+1} \leq \frac{2}{r_j} \leq \frac{2}{\sqrt{q_{j-1}}-1} \,.$$

We now replace each $\omega$ by its rational approximation $m + p_j/q_j$. It then follows easily since

$$|x_{sq_j+l}| \leq (r_j+1)q_j$$

and since for real numbers $a$ and $b$

$$\left| e^{ia} - e^{ib} \right| \leq |a - b| \,,$$

that the total error using the rational approximation of $\omega$ is bounded by

$$\left| \frac{1}{(2r_j+1)q_j} \sum_{s=-r_j, l=0}^{s=r_j, l=q_j-1} \left( e^{2\pi i \omega x_{sq_j+l}} - e^{2\pi i (m+p_j/q_j) x_{sq_j+l}} \right) \right| \leq 2\pi \frac{(r_j+1)q_j}{q_j q_{j+1}} \leq \frac{4\pi}{\sqrt{q_j}} \,.$$

For any given $(x_n) \in \mathcal{B}$, we have by definition $(x_n) \in \mathcal{B}(k, m)$, and therefore by Corollary 4 there is an integer $\xi$ such that if $r \geq \xi$ we have for any $j$

$$\left| \frac{1}{(2r+1)q_j} \sum_{s=-r, l=0}^{s=r, l=q_j-1} e^{2\pi i (m+p_j/q_j) x_{sq_j+l}} \right| \leq \frac{1}{k} \,.$$

The result follows at once for $\omega$ irrational since if $L \to \infty$, we have $q_{j-1} \to \infty$, which implies $r_j \to \infty$. If $\omega$ is rational (but not zero) we have $\omega = m + p/q$ with $m, p, q \in \mathbf{N}$, and $p \neq 0$ if $m = 0$. We then define $r \in \mathbf{N}$ and $t \in \mathbf{N}$ uniquely by $L = rq + t$ with $0 \leq t < q$ (if $p = 0$ we choose $q = 1$ and $r = L$). We then proceed as before (without of course making any rational approximation of $\omega$). The last statement of the proposition is now obvious.

**Proof.** (of Theorem 1) Let $f$ be a complex valued almost periodic function. For any number $\epsilon > 0$, there is a trigonometric polynomial $g$ approximating $f$ uniformly (see [B.] or [F.]), namely there are $a$ frequencies $\omega_1, \cdots, \omega_a$ and $a$ complex numbers $B_1, \cdots, B_a$, depending on $\epsilon$ such that if

$$g(x) = \sum_{k=1}^{a} B_k e^{2\pi i \omega_k x} ,$$

then

$$\sup_{x \in \mathbf{R}} |f(x) - g(x)| \leq \epsilon .$$

From the hypothesis on the zeroes of $f$ we deduce that for any $\omega \in \mathbf{R}$ and $(x_n) \in \mathcal{B}$ we have

$$\left| \lim_{L \to \infty} \frac{1}{(2L+1)} \sum_{-L \leq n \leq L} e^{-2\pi i \omega x_n} g(x_n) \right| = \left| \lim_{L \to \infty} \frac{1}{(2L+1)} \sum_{-L \leq n \leq L} e^{-2\pi i \omega x_n} (g(x_n) - f(x_n)) \right| \leq \epsilon .$$

Using Proposition 5, it follows that

$$\sup_{k} |B_k| \leq \epsilon.$$

From this we now conclude (using again the uniform approximation of $f$ by $g$) that for any $\omega$ we have

$$\left| \lim_{L \to \infty} \frac{1}{2L} \int_{-L}^{L} e^{-2\pi i \omega x} f(x) \, dx \right| \leq 2\epsilon .$$

Since the result is true for any $\epsilon > 0$, we conclude that $f$ has a Fourier spectrum (in the sense of almost periodic functions) equal to zero and this implies (see [B.], [F.]) that $f$ is the zero function.

## III. PROPERTIES OF A RANDOM BLASHKE PRODUCT.

We now start the proof of Theorem 2. In order to prove that $B$ is in $S'$ we have to prove that the following limit exists

$$\lim_{L\to\infty} \frac{1}{2L} \int_{-L}^{L} B(x+y)B(y)dy$$

for any $x$ and defines a continuous function. Since $B$ is analytic and bounded in a strip around the real axis, it is easy to verify that it is a uniformly Lipschitz function. It follows easily from this observation that if the above limit exists for every rational $x$ it will exist for all the reals and will be a Lipschitz function (in fact it will extend to a bounded analytic function in a strip). We will therefore prove below the existence of the limit only for $x$ rational.

We can also write for a fixed $x$

$$\int_{-L}^{L} B(x+y)B(y)dy = \sum_{j=-L}^{L-1} \xi_j$$

where

$$\xi_j = \int_0^1 B(x+y+j)B(y+j)dy \ .$$

The random variables $\xi_j$ are not independent but are asymtotically independent. Namely we have the following result.

**Lemma 6.** *The number $\mu = \mathbb{E}(\xi_j)$ is independent of $j$ and there exists two constants $\eta > 0$ and $\sigma > 0$ such that for any $n$ and $m$ in $\mathbf{Z}$*

$$|\mathbb{E}\left((\xi_n - \mu)(\xi_m - \mu)\right)| \leq \eta e^{-\sigma|n-m|} \ .$$

**Proof.**

We have by definition of $B$

$$B(x+y+j)B(y+j) = \prod_{n\in\mathbf{Z}} \tanh(x+y+j-n-X_n)\tanh(y+j-n-X_n)$$

$$= \prod_{k\in\mathbf{Z}} \tanh(x+y-k-X_{k+j})\tanh(y-k-X_{k+j}) \ .$$

Therefore, since the random variables $(X_j)$ are independent and identically distributed, we have

$$\mathbb{E}(B(x+y+j)B(y+j)) = \prod_{k\in\mathbf{Z}} \mathbb{E}\left(\tanh(x+y-k-X)\tanh(y-k-X)\right) \ ,$$

and the first result follows, namely $\mu$ is independent of $j$.

In order to prove the second result we observe that for any interval $[-A, A]$, there are two constants $\rho_A > 0$ and $\gamma_A > 0$ such that for any integer $p$ and any $t \in [-A, A]$ we have for any configuration $(X_j)$

$$\left| 1 - \prod_{|j|>p} \text{Sign}(-j - X_j) \tanh(t - j - X_j) \right| \leq \rho_A e^{-\gamma_A p}.$$

This follows at once from $|\tanh(s)| = 1 - \mathcal{O}(e^{-2s})$ for large $s$. Therefore, if $x \in [-A+1, A-1]$, we have for $r = [|n-m|/2]$

$$\left| \mathbb{E}(\xi_n \xi_m) - \mathbb{E}\left( \int_0^1 dy \prod_{|k|<r} \tanh(x + y - k - X_{k+m}) \tanh(y - k - X_{k+m}) \right) \right.$$

$$\left. \times \mathbb{E}\left( \int_0^1 dz \prod_{|l|<r} \tanh(x + z - l - X_{l+n}) \tanh(z - l - X_{l+n}) \right) \right| \leq 2\rho_A e^{-\gamma_A r}.$$

By a similar estimate one has

$$\left| \mu^2 - \mathbb{E}\left( \int_0^1 dy \prod_{|k|<r} \tanh(x + y - k - X_{k+m}) \tanh(y - k - X_{k+m}) \right) \right.$$

$$\left. \times \mathbb{E}\left( \int_0^1 dz \prod_{|l|<r} \tanh(x + z - l - X_{l+n}) \tanh(z - l - X_{l+n}) \right) \right| \leq 2\rho_A e^{-\gamma_A r},$$

and the Lemma follows.

**Proof.** (of Theorem 2) Let $S_L$ denote the random variable

$$S_L = \sum_{j=-L}^{L-1} \xi_j.$$

We have from the previous proposition

$$\mathbb{E}\left(\frac{S_L}{2L}\right) = \mu$$

and

$$\mathbb{E}\left[\left(\frac{S_L}{2L} - \mu\right)^2\right] \leq \mathcal{O}(1) L^{-1}.$$

Therefore from Tchebychev's inequality, we have

$$\mathbb{P}\left(\left|\frac{S_L}{2L} - \mu\right| > s\right) \leq \mathcal{O}(1)\frac{1}{Ls^2} \,.$$

We now consider the sequences $L_k = k^4$ and $s_k = k^{-1}$. Since the sequence $1/(L_k s_k^2)$ is summable, we deduce that with probability one,

$$\lim_{k \to \infty} \frac{S_{L_k}}{2L_k} = \mu \,.$$

Moreover, if for any $L > 1$ we define $k$ uniquely by $L_k \leq L < L_{k+1}$, we have since the random variables $\xi_j$ are bounded by one

$$\left|\frac{S_L}{2L} - \frac{S_{L_k}}{2L_k}\right| \leq \frac{4(L - L_k)}{(2L_k)} \leq \frac{4(L_{k+1} - L_k)}{(2L_k)} \leq \mathcal{O}(1)k^{-1} \leq \mathcal{O}(1)L^{-1/4}$$

and for a fixed rational $x$ we have almost surely

$$\mu = \lim_{L \to \infty} \frac{S_L}{2L} = \lim_{L \to \infty} \frac{1}{2L} \int_{-L}^{L} B(x+y)B(y)dy \,.$$

We also have convergence almost surely for any rational $x$, and as explained above since the last expression is uniformly Lipschitz in $x$ we have convergence for all $x$.

An interesting fact about the (random) function $B(x)$ is that for almost every realization $(x_n)$, the correlation function $C_{B,B}(x)$ is independent of $(x_n)$ as we proved above, but also it converges to a periodic function when $x$ tends to infinity. To see this, let $u$ be the function defined by

$$u(x) = \prod_k \mathbb{E}\big(\tanh(x - k - X)\mathrm{Sign}(-k - X)\big) \,.$$

Using arguments similar to those in the proof of the Lemma 6, it is easy to verify that for large $x$

$$\mu(x) - \int_0^1 u(x+y)u(y)\,dy = o(1/x) \,.$$

Moreover, we have

$$u(x+1) = \prod_l \mathbb{E}\left(\tanh(x - l - X)\mathrm{Sign}(-l - X)\frac{\mathrm{Sign}(-l - 1 - X)}{\mathrm{Sign}(-l - X)}\right) \,.$$

However, since with probability one, $X \in ]0, 1[$, we have

$$\begin{cases} \frac{\mathrm{Sign}(-l-1-X)}{\mathrm{Sign}(-l-X)} = 1 & \text{if } l \notin \{1, 2\} \\ \frac{\mathrm{Sign}(-l-1-X)}{\mathrm{Sign}(-l-X)} = -1 & \text{if } l = 1 \text{ or } 2. \end{cases}$$

This implies $u(x+1) = u(x)$ and the asymptotic periodicity of $C_{B,B}(x)$.

## IV SOME EXTENSIONS.

In this section we will discuss some extensions of the previous results in particular to other point processes.

As mentioned in the introduction, Theorem 1 can be extended to any dimension, with essentially the same proof. We leave the details to the interested reader.

We now define another point process $\mathcal{P}'$ on the real line as follows. For a fixed scale $l$, we choose a lattice $\mathcal{L}$ of lattice size $l$. We now choose a positive number $0 < a < l/2$, and in each interval of size $2a$ centered at a point of the lattice we choose at random a point uniformly distributed in that interval and independently for the different points of the lattice. In other words, if $l = 1$ and the lattice is centered at the origin, we choose $0 < a < 1/2$, and a family of independent random variables variables $(X_n)_{n \in \mathbf{Z}}$ uniformly distributed on $[-a, a]$. Our point process $\mathcal{P}'$ is defined by

$$x_n = n + X_n .$$

**Theorem 7.** *With probability one (with respect to the point process $(x_n)$), any almost periodic function $f$, real analytic and such that*

$$f(x_n) = 0 \qquad \forall n \in \mathbf{Z}$$

*is identically zero.*

**Proof.** The proof goes along the same lines as the proof of Theorem 1, and we will only mention the differences. As before, by translation invariance and scale invariance we can assume that $l = 1$ and the lattice is centered at the origin. Lemma 3 is valid as before except for $p = 0$, and we deduce Proposition 5 in the case where $\omega \notin \mathbf{Z}$. We conclude as in the proof of Theorem 1 that $f$ must be a periodic function of period 1. However, from the independence of the $X_n$, it is easy to verify that with probability one, the set of numbers $\{x_n \pmod 1\}$ has an infinite (countable) cardinality. Therefore, using the periodicity, $f$ has an infinite number of different zeros on the interval $[0, 1]$. Since it is real analytic it must be the constant zero function.

The analog of Theorem 2 also holds with an identical proof.

We now consider the case of Poisson point processes. We recall that a Poisson point process with intensity $\lambda$ is a point process $(x_n)$ such that the distances between neighboring points are independent random variables with an exponential distribution of average $\lambda$. In other words, fixing an origin, we have for $n > 0$

$$x_n = \sum_{j=1}^{n} y_j$$

where the $(y_j)$ are non negative i.i.d random variables such that

$$\mathbb{P}(y_j > s) = e^{-s/\lambda} .$$

A similar statement holds for $n < 0$.

**Theorem 8.** *With probability one (with respect to the Poisson point process $(x_n)$ with intensity $\lambda$), any almost periodic function $f$ satisfying*

$$f(x_n) = 0 \quad \forall n \in \mathbf{Z}$$

*is identically zero.*

Note that contrary to the previous case we do not require here $f$ to be real analytic.

As before, using scale invariance it is enough to prove Theorem 8 for $\lambda = 1$ and we will make this assumption form now on. We will first have to modify Lemma 3 as follows.

**Lemma 9.** *There is a positive constant $C > 1$ such that for any $k \in \mathbf{N}$, any $m \in \mathbf{Z}$, any $p, q, r \in \mathbf{N}$ such that $0 \leq p < q$, $p/q \geq 1/2$ if $m = 0$ and $p/q \leq 1/2$ if $m = -1$, the set*

$$\mathcal{A}(k, m, q, r, p) = \left\{ \left| \frac{1}{(2r+1)q} \sum_{s=-r}^{r} \sum_{l=0}^{q-1} e^{2\pi i (m+p/q) x_{sq+l}} \right| \geq \frac{1}{k} \right\}$$

*satistfies*

$$\mathbb{P}(\mathcal{A}(k, m, q, r, p)) \leq 4 e^{-\frac{1}{Ck^2}(2r+1)q} .$$

*For $m = 0$, $p/q < 1/2$, $rp^2/q > 2$, we have for any $k \in \mathbf{N}$*

$$\mathbb{P}(\mathcal{A}(k, 0, q, r, p)) \leq e^{-rqp^4/k^2 C q^4},$$

*and for $m = -1$, $p/q > 1/2$, $r(q-p)^2/q > 2$, we have*

$$\mathbb{P}(\mathcal{A}(k, -1, q, r, p)) \leq e^{-rq(q-p)^4/k^2 C q^4} .$$

**Proof.** As in the proof of Lemma 3, we introduce the random variables

$$A^+(m, q, r, p) = \sum_{s=0}^{s=r} \sum_{l=0}^{l=q-1} \cos\left(2\pi \Omega x_{sq+l+1}\right)$$

$$B^+(m, q, r, p) = \sum_{s=0}^{s=r} \sum_{l=0}^{l=q-1} \sin\left(2\pi \Omega x_{sq+l+1}\right) ,$$

$$A^-(m, q, r, p) = \sum_{s=-r}^{s=-1} \sum_{l=0}^{l=q-1} \cos\left(2\pi \Omega x_{sq+l}\right)$$

$$B^-(m, q, r, p) = \sum_{s=-r}^{s=-1} \sum_{l=0}^{l=q-1} \sin\left(2\pi \Omega x_{sq+l}\right) ,$$

where $\Omega = m + p/q$.

Here we distinguish in the estimate the random variables with positive and negative indices $n$. For a fixed number $\sigma$ small enough we are going to estimate recursively (on $j$) the quantity

$$Z_j(m,q,p,\sigma) = \mathbb{E}\left(e^{\sigma \sum_{n=0}^{j-1} \cos(2\pi\Omega x_n) + a_j \cos(2\pi\Omega x_j) + b_j \sin(2\pi\Omega x_j)}\right),$$

where the numbers $a_j$ and $b_j$ are going to be chosen adequately below, the goal beeing of course to obtain an estimate for $j = (r+1)q$, $b_{(r+1)q} = 0$ and $a_{(r+1)q} = \sigma$. Using (for $j > 1$) $x_j = x_{j-1} + y_j$, we have

$$\mathbb{E}\left(e^{\sigma \sum_{n=0}^{j-1} \cos(2\pi\Omega x_n) + a_j \cos(2\pi\Omega x_j) + b_j \sin(2\pi\Omega x_j)}\right) = \mathbb{E}\left(e^{\sigma \sum_{n=0}^{j-1} \cos(2\pi\Omega x_n)}\right.$$

$$\left. e^{\cos(2\pi\Omega y_j)(a_j \cos(2\pi\Omega x_{j-1}) + b_j \sin(2\pi\Omega x_{j-1})) + \sin(2\pi\Omega y_j)(b_j \cos(2\pi\Omega x_{j-1}) - a_j \sin(2\pi\Omega x_{j-1}))}\right)$$

and assuming that
$$|a_j| + |b_j| \leq 1/2$$

we obtain using similar estimates as in the proof of Lemma 3 and the independence of $y_j$ and $x_k$ for $k < j$

$$Z_j(m,q,p,\sigma) \leq Z_{j-1}(m,q,p,\sigma)\, e^{8(a_j^2 + b_j^2)},$$

with
$$a_{j-1} = \frac{a_j}{1+\Omega^2} + \frac{b_j \Omega}{1+\Omega^2} + \sigma$$
$$b_{j-1} = -\frac{a_j \Omega}{1+\Omega^2} + \frac{b_j}{1+\Omega^2}.$$

In order to control this recursion, it is natural to define the sequence of positive numbers

$$\rho_j = \sqrt{a_j^2 + b_j^2}.$$

We have easily
$$\rho_{j-1} \leq |\sigma| + \frac{\rho_j}{\sqrt{1+\Omega^2}}.$$

Therefore, with the conditions $b_{(r+1)q-1} = 0$ and $a_{(r+1)q-1} = \sigma$ we obtain for $\sigma(1+\Omega^2)/\Omega^2 < 1/8$

$$\sup_{1 \leq j \leq (r+1)q} \rho_j \leq \sigma \frac{\sqrt{1+\Omega^2}}{\sqrt{1+\Omega^2} - 1} \leq \frac{2\sigma(1+\Omega^2)}{\Omega^2} \leq \frac{1}{2\sqrt{2}}.$$

This implies finally that there is a uniform positive constant $c$ such that if $|\Omega| \geq 1/2$ and $\sigma < 1/64$

$$Z_{(r+1)q}(m,q,p,\sigma) \leq e^{c\sigma}e^{crq\sigma^2},$$

and also if $p/q < 1/2$ and $\sigma < \Omega^2/16$

$$Z_{(r+1)q}(0,q,p,\sigma) \leq e^{cq^2p^{-2}\sigma}e^{crq^5p^{-4}\sigma^2},$$

and if $p/q > 1/2$ (and $\sigma < \Omega^2/16$)

$$Z_{(r+1)q}(-1,q,p,\sigma) \leq e^{cq^2(q-p)^{-2}\sigma}e^{crq^5(q-p)^{-4}\sigma^2}.$$

If $|\Omega| \geq 1/2$, we choose $\sigma = \pm\alpha k^{-1}$ with $\alpha$ a small enough positive number uniform in $m$ $r$, $q$, $p$, $k$. The proof proceeds as in Lemma 3. And we have analogous estimates for $A^-$ and $B^{\pm}$.

If $|\Omega| < 1/2$, $m = 0$, and $rq\Omega^2 > 2$ we choose $\sigma = p^4/(4cq^4k)$ and we obtain

$$\mathbb{P}(\mathcal{A}(k,0,q,r,p)) \leq e^{-rqp^4/16kcq^4}.$$

If $|\Omega| < 1/2$, $m = -1$, and $rq\Omega^2 > 2$ we choose $\sigma = (q-p)^4/(4cq^4k)$ and we obtain

$$\mathbb{P}(\mathcal{A}(k,0,q,r,p)) \leq e^{-rq(q-p)^4/16kcq^4}.$$

In order to derive a result analogous to Corollary 4, we first introduce two sequences of sets which will allow us to refine the estimates for $m = 0$ and $m = 1$. For any integer $v > 1$, we define

$$\mathcal{C}^+(k,v) = \left\{ \begin{array}{c} \exists\, \xi \in \mathbf{N} \text{ such that } \forall r \in \mathbf{N} \text{ with } r > \xi,\ \forall q \in \mathbf{N}\ \forall p \in \mathbf{N} \\ \text{satisfying } rp^2 > 2q \text{ and } 1/(v+1) \leq p/q < 1/v \\ \left| \dfrac{1}{(2r+1)q} \displaystyle\sum_{s=-r,l=0}^{s=r,l=q-1} e^{2\pi i(p/q)x_{sq+l}} \right| \leq \dfrac{1}{k} \end{array} \right\},$$

and

$$\mathcal{C}^-(k,v) = \left\{ \begin{array}{c} \exists\, \xi \in \mathbf{N} \text{ such that } \forall r \in \mathbf{N} \text{ with } r > \xi,\ \forall q \in \mathbf{N}\ \forall p \in \mathbf{N} \\ \text{satisfying } r(q-p)^2 > 2q \text{ and } 1/(v+1) \leq (q-p)/q < 1/v \\ \left| \dfrac{1}{(2r+1)q} \displaystyle\sum_{s=-r,l=0}^{s=r,l=q-1} e^{2\pi i((p-q)/q)x_{sq+l}} \right| \leq \dfrac{1}{k} \end{array} \right\}.$$

We also redefine the sets $\mathcal{B}(k,0)$ and $\mathcal{B}(k,-1)$ by

$$\mathcal{B}(k,0) = \left\{ \begin{array}{c} \exists\, \xi \in \mathbf{N} \text{ such that } \forall r \in \mathbf{N} \text{ with } r > \xi,\ \forall q \in \mathbf{N} \text{ and } 0 \leq p < q/2 \\ \left| \dfrac{1}{(2r+1)q} \displaystyle\sum_{s=-r,l=0}^{s=r,l=q-1} e^{2\pi i(p/q)x_{sq+l}} \right| \leq \dfrac{1}{k} \end{array} \right\},$$

and similarly we redefine

$$\mathcal{B}(k,-1) = \left\{ \begin{array}{c} \exists\, \xi \in \mathbf{N} \text{ such that } \forall r \in \mathbf{N} \text{ with } r > \xi\,,\ \forall q \in \mathbf{N} \text{ and } q/2 < p < q \\ \left| \dfrac{1}{(2r+1)q} \sum_{s=-r,l=0}^{s=r,l=q-1} e^{2\pi i((q-p)/q)x_{sq+l}} \right| \leq \dfrac{1}{k} \end{array} \right\}.$$

**Corollary 10.** *The set*

$$\mathcal{C} = \bigcap_{k \in \mathbf{N}} \bigcap_{m \in \mathbf{Z}} \mathcal{B}(k,m) \bigcap_{k \in \mathbf{N}} \bigcap_{v \in \mathbf{N}} \mathcal{C}^+(k,v) \bigcap_{k \in \mathbf{N}} \bigcap_{v \in \mathbf{N}} \mathcal{C}^-(k,v)$$

*has probability 1.*

The proof follows immediatly from Lemma 9 as in Corollary 4.

We will also need the following large deviation result to obtain the analog of Proposition 5.

**Lemma 11.** *For a Poisson point process $(x_n)$ with intensity one starting at the origin, we have for any $L \geq 1$*

$$\mathbb{P}\left( \sup_{1 \leq |n| \leq L} |x_n| > 2L \right) \leq 2e^{-L(1-\log 2)}\,.$$

*In particular, with probability one, there is a finite integer $L_0((x_n))$ such that for any $L > L_0((x_n))$ we have*

$$\sup_{1 \leq |n| \leq L} |x_n| < 2L\,.$$

**Proof.** Since for $n > 0$ the sequence $x_n$ is non decreasing, it is enough to estimate the probability of the event $\{x_L > 2L\}$. A similar argument holds for $n < 0$ and $x_{-L}$. Since $x_L$ is a sum of $L$ independent exponential random variables, we have

$$\mathbb{E}\left( e^{x_L/2} \right) = 2^L\,.$$

On the other hand, using Tchebichev's inequality, we have

$$e^L \mathbb{P}(x_L > 2L) \leq \mathbb{E}\left( e^{x_L/2} \right) = 2^L\,,$$

and the result follows.

We now give a proof of Proposition 5 using Corollary 10 instead of Corrllary 4. The first part of the proof, namely the choice of $j$ and the replacement of $L$ by $r_j q_j$ is unchanged. The only difference is when we replace $\omega$ by it's rational approximation. Then the error is bounded by $\sup_{|n| < (r_j+1)q_j} |x_n|$, and with probability one, if we take $j$ large enough, this quantity is bounded by $2(r_j+1)q_j$ by Lemma 11. The rest of the proof now proceeds as before, noting that if $\omega \in\, ]-1/2, 1/2[$ ($\omega \neq 0$), there is an integer $v$ such that for $j$ large enough, $1/(v+1) \leq p_j/q_j \leq 1/v$ if $\omega > 0$, and $1/(v+1) \leq (q_j - p_j)/q_j \leq 1/v$ if $\omega < 0$.

The proof of Theorem 8 is then similar to the proof of Theorem 1.

In order to derive an analog of Theorem 2 for a Poisson point process, one needs the following result.

**Lemma 11.** *For a Poisson point process $(x_n)$ with intensity one starting at the origin, we have for any $L > 0$*

$$\mathbb{P}(x_L \leq L/2) \leq e^{-L(\log 2 - 1/2)} \ .$$

**Proof.** This is again a large deviation estimate. We have

$$\mathbb{E}\left(e^{L-x_L}\right) = e^{L(1-\log 2)} \ ,$$

and the result follows by Tchebichev's inequality as in Lemma 10.

The rest of the proof of the analog of Theorem 2 proceeds as before, and we leave the details to thge interested reader.

*Acknowledgements.* Part of this work was done during a stay at the Mittag-Leffler Institute. The author thanks the IML for its hospitality and pleasant environment, and A. Kupiainen for the invitation to the workshop on Nonlinear PDE's, Turbulence, and Statistical Mechanics. The author also thanks M. Benedicks for many helpfull discussions and the Mathematics Department of the Kungliga Tekniska Högskolan for their kind hospitality and financial support.

**References.**

- [B.] A.Besicovitch. *Almost Periodic Functions.* Cambridge University Press, Cambridge 1932.
- [C.1] Thermodynamic limit of the Ginzburg-Landau equation. Nonlinearity **7**, 1175-1190 (1994).
- [C.2] Non linear parabolic evolutions in unbounded domains. In *Dynamics, Bifurcations and Symmetries*, pp 97-104, P.Chossat editor. Nato ASI 437,Plenum, New York, London 1994.
- [C.E.] Space-Time Behavior in Problems of Hydrodynamic Type: A Case Study (with J.-P.Eckmann). Nonlinearity **5**, 1265-1302 (1992).
- [C.G.T.] B.Cockburn et al. Determining degrees of freedom for nonlinear dissipative equations. Preprint (1995).
- [C.T.] P.Collet, E.Titi, in preparation.
- [F.] J.Favard. *Leçons Sur Les Fonctions Presque-Périodiques.* Gauthier-Villars, Paris 1933.
- [G.] J.Garnett. *Bounded Analytic Functions.* Academic Press, New York, 1981.
- [H.] K.Hoffman. Banach Spaces of Analytic Functions. Dover, New York 1968.
- [H.R.] G.Hardy, E.Right. *An Introduction to the Theory of Numbers.* Oxford Univ. Press, Oxford 1961.
- [T.B.D.vH.T] P.Takac et al. Analyticity of essentially bounded solutions to semi linear parabolic systems and validity of the Ginzburg-Landau equation. Siam J. Math. Anal. (to appear).
- [K.S.M.] P.Kirrmann, G.Schneider, A.Mielke. The validity of modulation equations for extended systems with cubic nonlinearities. Proc. Roy. Soc. Ed. **122A**, 85-91 (1992).

[R.B.S.] B.W.Roberts, E.Bodenschatz, J.P.Sethna. Defect–Defect Correlation functions, generic scale invariance, and the complex Ginzburg-Landau equation. Preprint, Cornell University CU-LASSP-BWR9501 (1995), patt-sol/9504001.

[V.H.] A.van Harten. On the validity of Ginzburg-Landau's equation. J. Nonlinear Science, **1**, 397-422 (1991).

[W.] N.Wiener. *The Fourier Integral and Certain of Its Applications*. Dover, New York 1958.